\documentstyle[12pt]{article}
\input{epsf}
\renewcommand{\theequation}{\arabic{section}.\arabic{equation}}
\newcommand{\beq}{\begin{equation}}
\newcommand{\eeq}{\end{equation}}
\newcommand{\be}{\begin{eqnarray}}
\newcommand{\ee}{\end{eqnarray}}
\newcommand{\bftau}{\mbox{\boldmath{$\tau$}}}
\newcommand{\partialslash}{\partial\hspace{-.5em}/\hspace{.15em}}
\begin{document}
\rightline{RUB-TPII-4/98}
\rightline{hep-ph/9804436}
\vspace{1cm}
\begin{center}
{\bf\LARGE Isovector unpolarized quark distribution in the nucleon
in the large--$N_c$ limit}
\\[1cm]
{\bf\large P.V.\ Pobylitsa}, {\bf\large M.V.\ Polyakov}
\\[0.2 cm]
{\it Petersburg Nuclear Physics Institute, Gatchina,
St.\ Petersburg 188350, Russia}
\\[0.3 cm]
{\bf\large K. Goeke}, {\bf\large T. Watabe}, {\bf\large and C.\ Weiss}
\\[0.2 cm]
{\it Institut f\"ur Theoretische Physik II,
Ruhr--Universit\"at Bochum, \\ D--44780 Bochum, Germany} \\
\end{center}
\vspace{1.5cm}
\begin{abstract}
\noindent
We calculate the isovector (flavor--nonsinglet) unpolarized quark-- and
antiquark distributions in the nucleon at a low normalization point in the
large--$N_c$ limit.  The nucleon is described as a soliton of the effective
chiral theory.  The isovector distribution appears in the next--to--leading
order of the $1/N_c$--expansion. Numerical results for the quark-- and
antiquark distributions compare well with the parametrizations of the data
at a low normalization point.  This large--$N_c$ approach gives a flavor
asymmetry of the antiquark distribution (violation of the Gottfried sum
rule) in good agreement with the measurements.
\end{abstract}
\vspace{1.5cm} PACS: 13.60.Hb, 14.20.Dh, 12.38.Lg, 12.39.Ki, 11.15.Pg  
\newpage
\tableofcontents
\newpage
\section{Introduction}
\setcounter{equation}{0}
\noindent
The parton distribution functions of the nucleon contain the
non-perturbative information which enters in the cross section for
deep--inelastic scattering and a variety of other hard processes. Their
scale dependence in the asymptotic region is governed by perturbative QCD
and well understood. The starting points of the perturbative evolution,
however, that is, the distributions at a relatively low normalization
point, belong to the domain of non-perturbative physics and can at present
only be estimated using approximate methods to deal with the problem of the
structure of the nucleon.
\par
A very useful approximation is the theoretical limit of a large number of
colors, $N_c \rightarrow \infty$. It is known that in this limit QCD
becomes equivalent to an effective theory of mesons, with baryons emerging
as solitonic excitations \cite{Witten}. At low energies the structure of
the effective theory is determined by the spontaneous breaking of chiral
symmetry.  The first realization of the idea of the nucleon as a soliton of
the pion field was proposed by Skyrme \cite{Skyrme62,ANW}, which was,
however, based on an arbitrary choice of higher--derivative terms of chiral
Lagrangian.  A more realistic effective action for the pion field is given
by the integral over quark fields with a dynamically generated mass,
interacting with the pion field in a minimal chirally invariant way
\cite{DE}. Such an effective action has been derived from the instanton
vacuum of QCD, which provides a microscopic mechanism for the dynamical
breaking of chiral symmetry \cite{DP86}.  The chiral quark--soliton model
of the nucleon based on this effective action \cite{DPP88,D98} has been
very successful in describing hadronic observables such as the nucleon
mass, $N\Delta$--splitting, electromagnetic and weak form factors {\it
etc.} \cite{Review}.
\par
Recently it has been demonstrated that it is possible to compute also the
leading--twist parton distributions at a low normalization point in the
chiral quark--soliton model of the nucleon
\cite{DPPPW96,DPPPW97,WG97}. This field--theoretic description of the
nucleon allows to preserve all general requirements on parton
distributions, such as positivity and the partonic sum rules which hold in
QCD. In particular, it allows a consistent calculation of the antiquark
distributions. The approach has by now been extended to transverse
polarized distributions \cite{PP96} as well as to off--forward quark
distributions \cite{PPPBGW97}.
\par
The large--$N_c$ limit implies a classification of the parton distribution
functions in ``large'' and ``small'' ones \cite{DPPPW96}.  Generally, at
large $N_c$ the quark distributions are concentrated at values of 
$x\sim 1/N_c$, and with the chiral quark--soliton model we aim to compute
them for values of $x$ of this parametric order. The distribution functions
which appear in the leading order of the $1/N_c$--expansion are of the form
\be 
D^{\rm large}(x) &\sim& N_c^2 \,\, \rho (N_c x) ,
\ee
where $\rho (y)$ is a stable function in the large $N_c$--limit, which
depends on the particular distribution considered. They are the isosinglet
unpolarized and isovector (longitudinally or transverse polarized)
distributions, which have been computed in
Refs.\cite{DPPPW96,DPPPW97,PP96}.  The isovector unpolarized and isosinglet
polarized distributions, on the other hand, appear only in the
next--to--leading order of the $1/N_c$--expansion and are of the form
\be
D^{\rm small}(x) &\sim& N_c \,\, \rho (N_c x) .
\ee
\par
In this paper we compute the isovector unpolarized quark-- and antiquark
distributions in the chiral quark--soliton model. The general method for
calculating the $1/N_c$--subleading distributions and a discussion of their
properties have been given in Ref.\cite{DPPPW96}. Here we actually compute
the distribution function, including the Gottfried sum, which is a measure
of the flavor asymmetry of the antiquark distribution at the low scale
\cite{Gottfried67,Kumano97}.  This task requires, among other things, to
generalize the analytical and numerical methods for the computation of
distribution functions \cite{DPPPW97} to the case of the
$1/N_c$--subleading distributions, taking into account the rotation of the
chiral soliton.
\par
The isovector distribution considered here is of particular interest for
understanding the relation of the low--scale parton distributions to the
chiral effective dynamics. The distribution functions computed within the
effective chiral theory correspond to distributions of ``constituent''
quarks and antiquarks, {\it i.e.}, objects which possess a structure in
terms of QCD quarks and gluon\footnote{For a more detailed discussion of
this complex issue we refer to Refs.\cite{PW97,BPW97}; see also
Ref.\cite{WG97}.} \cite{DPPPW96}. The parameter governing the compositeness
of the constituent quark is the ratio of the dynamical quark mass to the
ultraviolet cutoff of the effective theory --- the ``size'' of the
constituent quark. In the instanton vacuum this ratio is proportional to
the small packing fraction of instantons, $(\bar\rho / R )^2$.  Before
comparing the quark-- and antiquark distributions computed in the effective
chiral theory with the parametrizations of the data at a low normalization
point \cite{GRV95} one should ``resolve'' the structure of the constituent
quark. From the singlet distribution, 
$u(x) + d(x) + \bar u(x) + \bar d(x)$, ``resolving'' the constituent quark
structure one recovers the true singlet quark and gluon distributions.
Phenomenologically one finds that gluons carry about 30\% of the nucleon
momentum at a normalization point of $\mu \approx 600\,{\rm MeV}$, so in
the singlet case the ``resolution'' of the constituent quark structure is a
rather sizable effect \cite{DPPPW96,DPPPW97}. In the isovector case
considered here the change of the distribution due to the ``resolution'' of
the constituent quark structure is expected to be less important, as this
distribution does not mix with the gluon distribution and its normalization
is scale independent (isospin sum rule). Consequently, the isovector
distribution calculated in the effective chiral theory can almost directly
be compared with the parametrizations of Ref.\cite{GRV95}, allowing one to
draw conclusions about the model dynamics even in the absence of a complete
understanding of the structure of the constituent quark.
\par
The plan of this paper is as follows.  Section \ref{sec_nucleon} contains a
brief exposition of the effective chiral theory and the description of the
nucleon as a chiral soliton, including the semiclassical quantization
procedure.  In Section \ref{sec_isovector_distribution} we derive the
expressions for the isovector quark distribution functions in the effective
chiral theory and discuss their properties. We first outline the expansion
in the soliton angular velocity ($1/N_c$--expansion) which is necessary for
computing the ``small'' ($1/N_c$--suppressed) distribution functions, in
particular the isovector unpolarized one. We then discuss the important
issue of ultraviolet regularization, which has been treated in detail in
Ref.\cite{DPPPW97}.  We show that the distributions obtained in our
approach satisfy the isospin sum rule. We also discuss the Gottfried sum.
In Section \ref{sec_numerical} we briefly describe the numerical technique
used for computation of the ``small'' $1/N_c$--suppressed distribution
functions. Similarly to the ``large'' $1/N_c$--leading distributions they
are computed as sums over quark single--particle levels in the background
pion field; however, now one is dealing with double sums over levels.  We
then discuss the numerical results and compare them to the parametrizations
of the data at a low normalization point \cite{GRV95}. Conclusions and an
outlook are given in Section \ref{sec_conclusions}.
\par
A calculation of the isovector unpolarized and isosinglet polarized
distributions following the approach of Refs.\cite{DPPPW96,DPPPW97} has
recently been performed by Wakamatsu and Kubota \cite{WK97}.  However,
these authors have neglected certain contributions to the distribution
function, which are important in particular at small values of $x$, as we
shall discuss below. Also, a calculation of the isovector unpolarized
structure function in a related approach has been reported in
Ref.\cite{WGR97}. In that calculation, however, only the contribution of
the so-called valence level is taken into account. This approximation leads
to a number of inconsistencies, as has been discussed in
Refs.\cite{DPPPW96,DPPPW97}.
\section{The nucleon as a chiral soliton}
\setcounter{equation}{0}
\label{sec_nucleon}
The starting point for the chiral quark--soliton model of the nucleon is
the effective action for the pion field, which is obtained by integrating
over quark fields in the background pion field \cite{DE,DP86},
\be
\exp\left( i S_{\rm eff}[U (x)] \right) &=&
\int D\psi D\bar\psi \; \exp\left[ i\int d^4 x\,
\bar\psi(i\partialslash - M U^{\gamma_5})\psi\right] .
\label{effective_action}
\ee
Here, $\psi$ is the fermion field, $M$ the dynamical quark mass, which is
due to the spontaneous breaking of chiral symmetry, and the pion (Goldstone
boson) field is described by an $SU(2)$ matrix, $U(x)$, with
\be
U^{\gamma_5}(x) &=& \frac{1+\gamma_5}2 U (x)
+ \frac{1-\gamma_5}2 U^\dagger (x) .
\label{U_gamma5}
\ee
In the long--wavelength limit, expanding in derivatives of the pion field,
the effective action Eq.(\ref{effective_action}) reproduces the
Gasser--Leutwyler Lagrangian with correct coefficients, including the
Wess--Zumino term. It is understood that the effective theory defined by
Eq.(\ref{effective_action}) is valid for momenta up to an UV cutoff, which
is the scale at which the dynamical quark mass drops to zero. We shall take
in the discussion here the quark mass to be momentum--independent and
assume divergent quantities to be made finite by applying some UV
regularization later. [Why this is generally justified will be discussed
below.]
\par
The effective action Eq.(\ref{effective_action}) has been derived from the
instanton vacuum, which provides a natural mechanism of dynamical chiral
symmetry breaking and enables one to express the parameters entering in
Eq.(\ref{effective_action}) --- the dynamical mass, $M$, and the
ultraviolet cutoff --- in terms of the QCD scale parameter, $\Lambda_{QCD}$
\cite{DP86}. In particular, the cutoff is given by the average instanton
size, $\bar\rho^{-1} \simeq 600 \, {\rm MeV}$
\par
In the effective chiral theory defined by Eq.(\ref{effective_action}) the
nucleon is in the large--$N_c$ limit characterized by a classical pion
field (``soliton''). In the nucleon rest frame it is of ``hedgehog'' form
\cite{DPP88},
\be
U_c ({\bf x}) &=& \exp\left[ i ({\bf n} \cdot \bftau ) P(r) \right] ,
\nonumber \\
r &=& |{\bf x}|, \hspace{1cm} {\bf n} \; = \; \frac{{\bf x}}{r} ,
\label{hedge}
\ee
where $P(r)$ is called the profile function, with $P(0) = -\pi$ and 
$P(r) \rightarrow 0$ for $r\rightarrow\infty$.  Quarks are described by
one-particle wave functions, which are determined as the solutions of the
Dirac equation in the background pion field,
\be
H(U_c) |n\rangle &=& E_n |n\rangle .
\label{eigen_n}
\ee
Here $H(U_c )$ is the single--particle Dirac Hamiltonian in the background
pion field given by Eq.(\ref{hedge}),
\be
H(U) &=& - i\gamma^0\gamma^k \partial_k + M\gamma^0 U^{\gamma_5} .
\label{H}
\ee
The spectrum of $H(U_c)$ includes a discrete bound--state level, whose
energy is denoted by $E_{\rm lev}$, as well as the positive and negative
Dirac continuum, polarized by the presence of the pion field.  The soliton
profile, $P(r)$, is determined by minimizing the static energy of the pion
field, which is given by the sum of the energy of the bound--state level
and the aggregate energy of the negative Dirac continuum, the energy of the
free Dirac continuum $(U = 1)$ subtracted \cite{DPP88},
\be
E_{\rm tot} &=& N_c \left[ 
\sum_{\scriptstyle n \atop \scriptstyle {\rm occup.}} E_n 
- \sum_{\scriptstyle n \atop \scriptstyle {\rm occup.}} E_n^{(0)} 
\right]
\nonumber \\
&=& N_c E_{\rm lev} \; + \; 
N_c \sum_{\scriptstyle n \atop \scriptstyle {\rm neg. cont.}} 
( E_n - E_n^{(0)} ) ,
\label{E_tot}
\ee
and in the leading order of the $1/N_c$--expansion the nucleon mass is
given simply by the value of the energy at the minimum,
\be
M_N &=& E_{\rm tot}\Biggr|_{U = U_c} .
\label{M_N_static}
\ee
The expression for the energy of the pion field, Eq.(\ref{E_tot}), contains
a logarithmic ultraviolet divergence due to the contribution of the Dirac
continuum and requires regularization.  In the calculation of the parton
distributions below we shall use a Pauli--Villars regularization (see
Subsection \ref{subsec_regularization}), so we regularize also the energy
using the Pauli--Villars method. Following Ref.\cite{WG97} we define
\be
E_{\rm tot, reg} &=&  N_c E_{\rm lev}
\; + \; N_c \left[ \sum_{\scriptstyle n \atop \scriptstyle {\rm neg. cont.}} 
( E_n - E_n^{(0)} )
\; - \; \frac{M^2}{M_{PV}^2}
\sum_{\scriptstyle n \atop \scriptstyle {\rm neg. cont.}} 
( E_{PV, n} - E_{PV, n}^{(0)} ) \right] ,
\label{E_tot_reg}
\ee
where $E_{PV, n}$ are the eigenvalues of the Hamiltonian, Eq.(\ref{H}),
with $M$ replaced by a regulator mass, $M_{PV}$, and the same pion
field. The value of the regulator mass, which now plays the role of the
physical cutoff, can be fixed from the pion decay constant,
\be
F_\pi^2 &=& \frac{N_c M^2}{4\pi^2} \log \frac{M_{PV}^2}{M^2} .
\label{fpi}
\ee
[Numerically, $F_\pi = 93\, {\rm MeV}$.] Note that we do not subtract the
finite contribution of the discrete level for the spectrum with the
Pauli--Villars mass; this prescription leads to a stable minimum of the
regularized energy functional, Eq.(\ref{E_tot_reg}), with respect to the
profile function \cite{WG97}. The soliton profile (the ``self-consistent''
pion field) for this UV regularization has been determined in
Refs.\cite{WG97,Doering92}.
\par
In higher order of the $1/N_c$--expansion one must take into account the
quantum fluctuations about the saddle--point pion field. A special role
play the zero modes of the pion field. In fact, the minimum of the energy,
Eq.(\ref{E_tot}), is degenerate with respect to translations of the soliton
field in space and to rotations in ordinary and isospin space. For the
hedgehog field, Eq.(\ref{hedge}), the two rotations are
equivalent. Quantizing the zero modes modes, {\it i.e.}, integrating over
collective rotations and translations, gives rise to nucleon states with
definite momentum and spin/isospin quantum numbers \cite{ANW,DPP88}. One
performs a (time--dependent) rotation of the hedgehog field,
Eq.(\ref{hedge}),
\be
U_c ({\bf x}) &\rightarrow& R(t) \, U_c({\bf x}) \, R^\dagger (t) ,
\label{rotation_U}
\ee
where the collective coordinate, $R(t)$, is a rotation matrix in
$SU(2)$--flavor space. The functional integral over the collective
coordinate can be computed systematically within the
$1/N_c$--expansion. The moment of inertia of the soliton is $O(N_c )$ (the
nucleon is ``heavy''), hence the angular velocity is
\be
\Omega &\equiv& \Omega_a \frac{\tau^a}{2}
\;\; = \;\; -i R^\dagger \dot{R} \;\; = \;\;
O\left( \frac{1}{N_c} \right)
\label{omega_def}
\ee
(the collective motion is ``slow''), and one can expand in powers of the
angular velocity. To leading order in $\Omega$ the collective motion is
described by a Hamiltonian
\be
H_{\rm rot} &=& \frac{S_a^2}{2 I} \;\; = \;\; \frac{T_a^2}{2 I} ,
\label{H_rotator}
\ee
where $S_a$ and $T_a$ are the right and left angular momenta, and the
Hamiltonian Eq.(\ref{H_rotator}) has been obtained by the ``quantization
rule''
\be
\Omega_a &\rightarrow& \frac{S_a}{I} .
\label{quantization_rule}
\ee
Here $I$ denotes the moment of inertia of the soliton. It can be expressed
as a double sum over quark single--particle levels in the background pion
field,
\be
I &=& \frac{N_c}{6} \sum_{\scriptstyle n \atop \scriptstyle {\rm occup.}}
\sum_{\scriptstyle m \atop \scriptstyle {\rm non-occup.}}
\frac{\langle n | \tau^a | m \rangle \langle m | \tau^a | n \rangle}
{E_m - E_n} .
\label{I}
\ee
Here the sum over $n$ runs over all occupied states, {\it i.e.}, the
discrete level and the negative Dirac continuum, the sum over $m$ over all
non-occupied states, {\it i.e.}, the positive Dirac continuum.
\par
The Hamiltonian Eq.(\ref{H_rotator}) describes a spherical top in
spin/isospin space, subject to the constraint $S^2 = T^2$, which is a
consequence of the ``hedgehog'' symmetry of the static pion field,
Eq.(\ref{hedge}). Its eigenfunctions, classified by $S^2 = T^2, S_3$ and
$T_3$ are given by the Wigner finite--rotation matrices \cite{DPP88},
\be
\phi^{S=T}_{S_3 T_3}(R) &=&
\sqrt{2S+1} (-1)^{T+T_3} D^{S=T}_{-T_3,S_3}(R) .
\label{Wigner}
\ee
The four nucleon states have $S = T = 1/2$, with $S_3, T_3 = \pm 1/2$,
while for $S = T = 3/2$ one obtains the 16 states of the $\Delta$
resonance.  The rotational energy, $S(S + 1)/(2 I)$, gives a
$1/N_c$--correction to the nucleon mass, which should be added to
Eq.(\ref{M_N_static}).  In particular, the nucleon--$\Delta$ mass splitting
is given by
\be
M_\Delta - M_N &=& \frac{3}{2 I} .
\ee
\par
The expression for the moment of inertia, Eq.(\ref{I}), contains an
ultraviolet divergence which is to be removed by the ultraviolet
cutoff. The ultraviolet regularization of the moment of inertia must be
consistent with that of the isovector quark distribution function; it will
be discussed below in Subsection \ref{subsec_regularization}.
\section{The isovector unpolarized quark distribution}
\setcounter{equation}{0}
\label{sec_isovector_distribution}
\subsection{The isovector distribution function in the large--$N_c$ limit}
\label{subsection_isovector_quark_distribution}
To compute the twist--2 quark and antiquark distribution functions in the
effective chiral theory we start from their ``field--theoretic'' definition
as forward matrix elements of certain light--ray operators in the nucleon,
which can be regarded as generating functions for the local twist--2
operators \cite{CollinsSoper82}.  Alternatively, one could start from the
``parton model'' definition as numbers of particles carrying a given
fraction of the nucleon momentum in the infinite--momentum frame
\cite{Feynman72} --- both ways lead to identical expressions for the quark
distribution functions in the chiral quark--soliton model
\cite{DPPPW96,DPPPW97}.  The unpolarized distribution ($f$ denotes the
quark flavor) is given by
\be
D_{{\rm unpol}, f} (x) &=&  \frac{1}{4\pi}
\int\limits_{-\infty}^\infty dz^- \, e^{ixp^+z^-} \,
\langle P | \, \bar\psi_f (0) \, \gamma^+ \, \psi_f (z) \, | P \rangle
\Bigr|_{z^+=0,\>z_\perp=0}\, , \nonumber \\
\bar D_{{\rm unpol}, f} (x) &=& - \left\{ x \rightarrow -x \right\} .
\label{nonlocal}
\ee
Here, $z^\pm$ and $\gamma^\pm$ denote the usual light--like vector
components and Dirac matrices,
\be
z^{\pm} &=& \frac{z^0\pm z^3}{\sqrt 2}, \hspace{1cm}
\gamma^{\pm} \;\; = \;\; \frac{\gamma^0 \pm \gamma^3}{\sqrt 2} .
\label{cone-coordinates}
\ee
In the longitudinally polarized case one should replace in
Eq.(\ref{nonlocal}) $\gamma^+ \rightarrow \gamma^+ \gamma_5$, and the
polarized antiquark distribution is given by the function at 
$x \rightarrow -x$ without minus sign. In Eq.(\ref{nonlocal}) we have
dropped the gauge field degrees of freedom (the path--ordered exponential
of the gauge field), a step which is justified when working in leading
order in the ratio of $M$ to the UV cutoff, {\it viz.}\ the packing
fraction of the instanton medium; see Refs.\cite{PW97,BPW97} for a detailed
discussion.
\par
Thanks to its relativistically invariant definition the matrix element
Eq.(\ref{nonlocal}) can be evaluated in any frame; for us it is convenient
to compute it in the nucleon rest frame.  The calculation follows the usual
procedure for computing matrix elements of quark bilinears in the chiral
quark--soliton model within the $1/N_c$ expansion. The limit 
$N_c \rightarrow \infty$ justifies the use of the saddle point
approximation for the pion field. The matrix element can be calculated with
the help of the quark Feynman Green function in the background pion field,
\be
G_F (y^0, {\bf y}; x^0, {\bf x})
&=&
\langle y^0, {\bf y} | \left[ i\partial_t - H(U) \right]^{-1}
| x^0, {\bf x} \rangle .
\label{G_F}
\ee
Here the saddle--point pion field is the slowly rotating hedgehog field,
Eq.(\ref{rotation_U}). For this ansatz the Green function, Eq.(\ref{G_F}),
takes the form
\be
\left[
i\partial_t - H(U) \right]^{-1} &=&
R(t) \; [i\partial_t - H(U_c) - \Omega ]^{-1} \; R^\dagger (t) ,
\label{rotH}
\ee
where $\Omega$ is the angular velocity, Eq.(\ref{omega_def}).  The matrix
element of a quark bilinear between nucleon states of given spin and flavor
quantum numbers is obtained by integrating over soliton rotations with wave
functions of the collective coordinates, $R$, corresponding to a given
spin/isospin state. In addition, one has to perform a shift of the center
of the soliton and integrate over it with plane--wave wave functions in
order to obtain a nucleon state of definite three--momentum.  This leads to
the following expression for the forward matrix element of a color--singlet
quark bilinear in the nucleon state (${\rm T}$ denotes the time--ordered
product):
\be
\lefteqn{
\langle {{\bf P}=0,S=T,S_3,T_3} | \,
{\rm T} \left\{ \psi^\dagger(x) \Gamma 
\psi(y) \right\} | {{\bf P}=0,S=T,S_3,T_3} \rangle }
\nonumber \\
&=& - 2 i M_N N_c
\int dR_1 \int dR_2 \left[\phi_{T_3S_3}^{T=S}(R_2)\right]^\ast
\phi_{T_3S_3}^{T=S}(R_1)
\int\limits_{R(-T)=R_1}^{R(T)=R_2} \! DR \;
\mbox{Det} \left[ i\partial_t - H(U_c) - \Omega \right]
\nonumber \\
&& \times
\int d^3{\bf X}  \;
\mbox{Tr} \left[ R^\dagger(x^0) \Gamma R(y^0) \;
\langle {y^0,{\bf y}-{\bf X}} | \;
\left[ i\partial_t - H(U_c) - \Omega \right]^{-1}
|{x^0,{\bf x}-{\bf X}} \; \rangle  \right] .
\label{me_rotation}
\ee
Here $\Gamma$ denotes a matrix in Dirac spinor and isospin space, and
$\mbox{Tr}\ldots$ implies the trace over Dirac and flavor indices (the sum
over color indices has already been performed).  The path integral over
$R(t)$ can be computed using the fact that in the large $N_c$ limit the
angular velocity of the soliton, $\Omega = - i R^\dagger \dot R$,
Eq.(\ref{omega_def}), is suppressed, which allows to expand both the Dirac
determinant and the propagator in the integrand in local powers of
derivatives of $R(t)$. In particular this expansion gives rise to the
kinetic term
\beq
\exp\left[ \frac{i}{2I}\; \int dt \; \Omega_a^2 (t)
\right] ,
\eeq
where $I$ is the moment of inertia of the soliton, Eq.(\ref{I}). The path
integral over $R(t)$ with this action can now be computed exactly; it
corresponds to the rigid rotator described by the Hamiltonian
Eq.(\ref{H_rotator}).
\par
When expanding the integrand of the path integral in Eq.(\ref{me_rotation})
in powers of $\Omega$ we keep all linear terms in $\Omega$ in addition to
the exponentiated kinetic term. The calculation of the path integral over
$R(t)$ is then equivalent to replacing the angular velocity by the spin
operator, $S$, according to the ``quantization rule'',
Eq.(\ref{quantization_rule}).  When dealing with the matrix element of a
non-local bilinear operator as in Eq.(\ref{me_rotation}), terms linear in
$\Omega$ in the integrand of the path integral arise from two sources:
\begin{itemize}
\item
Expansion of the quark propagator ({\it cf.} Eq.(\ref{rotH})):
\be
\lefteqn{ [i\partial_t - H(U_c) - \Omega ] }
&& \nonumber \\
&=& [i\partial_t - H(U_c)]^{-1} 
+ [i\partial_t - H(U_c)]^{-1} \; \Omega \;
[i\partial_t - H(U_c)]^{-1} + \ldots
\label{propagator-expansion}
\ee
\item
Expansion of the nonlocal object $R^\dagger (x^0 ) \Gamma R (y^0 )$.  This
expansion can be performed in two alternative ways:
\be
R^\dagger(x^0) \Gamma R(y^0)
&=& R^\dagger(x^0) \Gamma R(x^0)
+ (y^0-x^0) R^\dagger(x^0) \Gamma \dot R(x^0) + \ldots
\label{choice-1}
\\
\mbox{or} \hspace{1cm} 
R^\dagger(x^0) \Gamma R(y^0)
&=& R^\dagger(y^0) \Gamma R(y^0)
- (y^0-x^0) \dot R{}^\dagger(y^0) \Gamma R(y^0) + \ldots
\ee
It can be shown \cite{DPPPW96} that both choices lead to the same result
for the rotational correction to the matrix element.
\end{itemize}
\par
Turning now to the calculation of parton distribution functions, defined as
matrix elements of the non-local operator Eq.(\ref{nonlocal}), it was shown
in Ref.\cite{DPPPW96} that the isosinglet unpolarized distribution, 
$u(x) + d(x)$, is non-zero already in the leading order of the
$1/N_c$--expansion, {\it i.e.}, at order $\Omega^0$ in the expansion of the
integrand of Eq.(\ref{me_rotation}). The isovector unpolarized
distribution, on the other hand, appears only in the next--to--leading
order, after expanding the integrand to order $\Omega^1$. It consists of
two pieces, which arise, respectively, from the expansion of the
propagator, Eq.(\ref{propagator-expansion}), and due to the non-locality of
the operator, {\it cf.}\ Eq.(\ref{choice-1}),
\be
u(x) - d(x) &=& [u(x) - d(x)]^{(1)} \; + \; [u(x) - d(x)]^{(2)} .
\label{isovec_1_2}
\ee
The first contribution from the expansion of the propagator,
Eq.(\ref{propagator-expansion}), is given by
\be
\lefteqn{ [u(x) - d(x)]^{(1)} \;\; = \;\;
-\frac{iM_N N_c}{4\pi} } && \nonumber \\
&&\times \sum\limits_{S_3}
\int\limits_{-\infty}^\infty dz^0  e^{ixM_Nz^0}
\int d^3{\bf X} \int dR \,
\left[\phi_{T_3S_3}^{T=S}(R)\right]^\ast \;
\mbox{Tr} \left[ \, R^\dagger \tau^3 R \;
(1+\gamma^0\gamma^3) \;
\right.
\nonumber \\
&&
\times
\left.
\langle {z^0,{\bf z}-{\bf X}} | \;
\frac1{i\partial_t - H(U_c)} \,
\left( \frac{1}{2I} S^a\tau^a \right)
\frac1{i\partial_t - H(U_c)} \;
|{0,-{\bf X}}\rangle  \right]
\Bigr|_{z^3=-z^0,\>z_\perp=0}
\phi_{T_3S_3}^{T=S}(R) .
\nonumber \\
\label{omegaexp_1}
\ee
The second contribution, originating from the non-locality of the operator,
Eq.(\ref{choice-1}), is given by
\be
\lefteqn{[u(x) - d(x)]^{(2)} \;\; = \;\;
-\frac{iM_N N_c}{4\pi} } 
\nonumber \\
&& \times \sum\limits_{S_3}
\int\limits_{-\infty}^\infty dz^0  e^{i x M_N z^0}
\int d^3{\bf X} \int dR \,
\left[\phi_{T_3S_3}^{T=S}(R)\right]^\ast 
\; \mbox{Tr} \left[ \, z^0 R^\dagger \tau^3 R \; \frac{i}{2I} S^a\tau^a \;
(1+\gamma^0\gamma^3) \right.
\nonumber \\
&& \times \left.
\langle{z^0,{\bf z}-{\bf X}}| 
\frac1{i\partial_t - H(U_c)} \;
|{0,-{\bf X}} \rangle  \right]
\Bigr|_{z^3=-z^0,\>z_\perp=0}
\phi_{T_3S_3}^{T=S}(R) .
\label{omegaexp_2}
\end{eqnarray}
Here $S^a$ is the spin operator, acting on the rotational wave functions,
which enters through the ``quantization rule'',
Eq.(\ref{quantization_rule}). Now it is a matter of routine calculations,
typical for the chiral quark--soliton model, to compute the rotational
matrix elements in Eqs.(\ref{omegaexp_1}, \ref{omegaexp_2}) and evaluate
the functional traces using the basis of eigenfunctions of the Dirac
Hamiltonian, Eq.(\ref{H}).  The details can be found in Appendix
\ref{sec_appendix}.  The result for the first contribution,
Eq.(\ref{omegaexp_1}), is
\be
\lefteqn{ [u(x)-d(x)]^{(1)} \;\;  = \;\; (2T^3) \frac{N_cM_N}{12 I} } &&
\nonumber \\
&\times& \left\{
2\sum\limits_{\scriptstyle n\atop \scriptstyle{\rm occup.}}
\sum\limits_{\scriptstyle m\atop {\scriptstyle {E_m \neq E_n}}}
\frac{1}{E_m - E_n}
\langle n|\tau ^a|m\rangle
\langle m|\tau ^a (1+ \gamma^0 \gamma^3 ) \delta (E_n+P^3-xM_N) |n \rangle
\right.
\nonumber \\
&& - \left. \sum\limits_{\scriptstyle n\atop \scriptstyle{\rm occup. }}
\sum\limits_{\scriptstyle m\atop {\scriptstyle {E_m = E_n}}}
\langle n|\tau^a|m\rangle
\langle m|\tau ^a(1+\gamma ^0\gamma ^3)
\delta ^{\prime }(E_n+P^3-xM_N)|n\rangle
\right\} ,
\label{isovec_1_occ}
\ee
where $T^3 = \pm 1/2$ is the isospin projection of the nucleon.
Eq.(\ref{isovec_1_occ}) expresses the isovector distribution as a sum over
quark single--particle levels in the background pion field, {\it cf.}\
Eq.(\ref{eigen_n}).  In the first term on the R.H.S.\ of
Eq.(\ref{isovec_1_occ}) the outer sum over $n$ runs over all occupied
levels, including the bound--state level as well as the negative Dirac
continuum, while the inner sum runs over {\it all} levels with the
restriction that $E_m \neq E_n$.\footnote{Here, as in Eqs.(\ref{E_tot},
\ref{I}), the sums over $n, m$ denote the sum over all quantum numbers
characterizing the single--particle states; note that the energy
eigenvalues are in general degenerate in the third component of the ``grand
spin'', the sum of the quark angular momentum and isospin \cite{DPP88}.}
This formula assumes a quasi--discrete spectrum and can be directly applied
to numerical calculations in a finite box; for the continuum case see
Appendix \ref{sec_appendix}.  The second contribution,
Eq.(\ref{omegaexp_2}), can be expressed through the derivative of the {\it
isosinglet} quark distribution, calculated in leading order of the
$1/N_c$--expansion:
\be
[u(x)-d(x)]^{(2)} &=&
-(2T^3) \frac{ N_c}{4I} \times \frac{\partial}{\partial x}
\sum\limits_{\scriptstyle n\atop \scriptstyle{\rm occup.}}
\langle n|  (1+\gamma^0\gamma^3)   \delta(E_n +P^3 - xM_N)   |{n}\rangle
\nonumber\\
&=& -(2T^3) \frac{1}{4IM_N} \times \frac{\partial}{\partial x}
[u(x)+d(x)]^{\rm leading} .
\label{isovec_2_occ}
\ee
\par
We note that, as in the case of the isosinglet distribution
\cite{DPPPW96,DPPPW97}, there exists an equivalent representation of the
isovector distribution as a sum over {\it non-occupied} levels (see
Appendix \ref{sec_appendix}):
\be
\lefteqn{ [u(x)-d(x)]^{(1)} \;\; = \;\;
-(2T^3) \frac{N_c M_N}{12 I} } &&
\nonumber \\
&\times& \left\{
2 \sum\limits_{\scriptstyle n\atop \scriptstyle{\rm non-occup.}}
\sum\limits_{\scriptstyle m\atop {\scriptstyle {E_m \neq E_n}}}
\frac{1}{E_m - E_n} 
\langle n|\tau ^a|m\rangle
\langle m|\tau ^a (1+ \gamma^0 \gamma^3 ) \delta (E_n+P^3-xM_N) |n \rangle
\right. 
\nonumber \\
&& - \left. \sum\limits_{\scriptstyle n\atop \scriptstyle{\rm non-occup. }}
\sum\limits_{\scriptstyle m \atop {\scriptstyle {E_m = E_n}}}
\langle n|\tau^a|m\rangle
\langle m|\tau ^a(1+\gamma ^0\gamma ^3) 
\delta ^{\prime }(E_n+P^3-xM_N)|n\rangle
\right\} ,
\label{isovec_1_nonocc}
\ee
\be
[u(x)-d(x)]^{(2)} &=&
(2T^3) \frac{ N_c}{4I} \times \frac{\partial}{\partial x}
\sum\limits_{\scriptstyle n\atop \scriptstyle{\rm non-occup.}}
\langle n|  (1+\gamma^0\gamma^3)   \delta(E_n +P^3 - xM_N)   |{n}\rangle .
\nonumber \\
\label{isovec_2_nonocc}
\ee
\par
In both Eqs.(\ref{isovec_1_occ}, \ref{isovec_2_occ}) and
(\ref{isovec_1_nonocc}, \ref{isovec_2_nonocc}) vacuum subtraction, {\it
i.e.}, subtraction of the corresponding sums over vacuum levels ($U = 1$)
and with vacuum occupation numbers, is understood.  It can be shown that
vacuum subtraction is required only for $x < 0$ in the sum over occupied,
and for $x > 0$ in the sum over non-occupied levels.  This fact is
important for the numerical calculations: we shall use the representation
as a sum over occupied states for $x > 0$, and as a sum over non-occupied
states for $x < 0$ (see below).
\par
The two contributions to the isovector distribution function,
Eq.(\ref{isovec_1_occ}) and Eq.(\ref{isovec_2_occ}), which have emerged
from the expansion of the integrand of Eq.(\ref{me_rotation}), both contain
delta function type singularities at $x = 0$. For the second contribution,
Eq.(\ref{isovec_2_occ}), this is immediately obvious: the isosinglet
distribution is discontinuous at $x = 0$, so its derivative contains a
delta function centered at $x = 0$. In a similar way one can convince
oneself that also the second term in Eq.(\ref{isovec_1_occ}) exhibits such
a singularity. These delta function singularities cancel when the two
contributions are added, {\it cf.}\ Eq.(\ref{isovec_1_2}). For the
numerical calculations it is convenient to regroup the terms in such a way
that this cancellation happens at the level of analytical expressions, {\it
i.e.}, that the singular terms are combined in one term. By inserting
intermediate states this term can be expressed as a double sum over levels,
similar to the first term on the L.H.S.\ of Eq.(\ref{isovec_1_occ}). Doing
this simple rearrangement one obtains
\be
[u(x)-d(x)] &=& [u(x)-d(x)]^{(1')} + [u(x)-d(x)]^{(2')} , 
\label{isovec_combined}
\ee
\be
\lefteqn{ [u(x)-d(x)]^{(1')} \;\; = \;\; 
(2T^3) \frac{N_cM_N}{6 I} } &&
\nonumber \\
&\times& 
\sum\limits_{\scriptstyle n\atop \scriptstyle{\rm occup.}}
\sum\limits_{\scriptstyle m\atop {\scriptstyle {E_m \neq E_n}}}
\frac{1}{E_m - E_n} 
\langle n|\tau ^a|m\rangle
\langle m|\tau ^a (1+ \gamma^0 \gamma^3 ) \delta (E_n+P^3-xM_N) |n \rangle ,
\label{isovec_combined_1}
\\
\lefteqn{[u(x)-d(x)]^{(2')} \;\; = \;\; - (2T^3) \frac{N_c}{12 I} } && 
\nonumber \\
&\times& \frac{\partial}{\partial x}
\sum\limits_{\scriptstyle n\atop \scriptstyle{\rm occup.}}
\sum\limits_{\scriptstyle m\atop \scriptstyle{E_m \neq E_n}}
\langle n| \tau^a |m \rangle 
\langle m| \tau^a (1+\gamma^0\gamma^3)   
\delta(E_n + P^3 - x M_N)| n \rangle ,
\label{isovec_combined_2} 
\ee
and a corresponding representation as sums over non-occupied states, {\it
cf.}\ Eqs.(\ref{isovec_1_nonocc}, \ref{isovec_2_nonocc}). The expressions
for the distribution function Eqs.(\ref{isovec_combined},
\ref{isovec_combined_1} \ref{isovec_combined_2}) will be used in the actual
numerical calculations.
\par
We would like to comment on a recent calculation of the isovector
distribution in the chiral quark--soliton model by Wakamatsu and Kubota
\cite{WK97}, who follow basically the same approach as we do here. However,
in the expansion of the integrand of the path integral over soliton
rotations, Eq.(\ref{me_rotation}), these authors drop the terms arising due
to the non-locality of the quark bilinear operator,
Eq.(\ref{isovec_combined_2}).  They argue that the non-locality of this
operator is of the order of the inverse large momentum transfer in
deep--inelastic scattering, $1/Q$, so that the slow rotational motion of
the soliton does not affect the operator. This argument, however, seems not
appropriate: the light--like separation between the quark fields in the
operator for the parton distribution is governed by the Bjorken variable
$x$ (more precisely, by $x M_N$), not by the momentum transfer $Q^2$. After
factorization of the DIS cross section in QCD the hard momentum enters only
in the coefficient functions, which can be computed perturbatively, not in
the operator for the parton distribution functions. Hence there is no
reason to drop the contribution Eq.(\ref{isovec_combined_2}) from the
$\Omega$--expansion, the more that the non-locality of the quark bilinear
operator is taken into account at other places in the calculation of
the parton distribution.
\par
The authors of Ref.\cite{WK97} also argue that in the contribution
Eq.(\ref{isovec_combined_1}) the double sum over levels should be
restricted to include only transitions from occupied to non-occupied
states, since transitions from occupied to occupied states, as are present
in Eq.(\ref{isovec_combined_1}) ({\it cf.}\ the derivation in Appendix
\ref{sec_appendix}) would violate the Pauli principle.  While this argument
is correct for local operators, we are dealing here with an operator
non-local in time, for which this so-called Pauli blocking is not
restrictive. We thus see no reason for dropping the occupied--to--occupied
transitions in the double sum over levels in
Eq.(\ref{isovec_combined_1}). Indeed, it turns out that the full expression
Eq.(\ref{isovec_combined_1}) is required to ensure equivalence of summation
over occupied and non-occupied states, as will be discussed below.
\par
\begin{figure}
\setlength{\epsfxsize}{15cm}
\setlength{\epsfysize}{15cm}
\epsffile{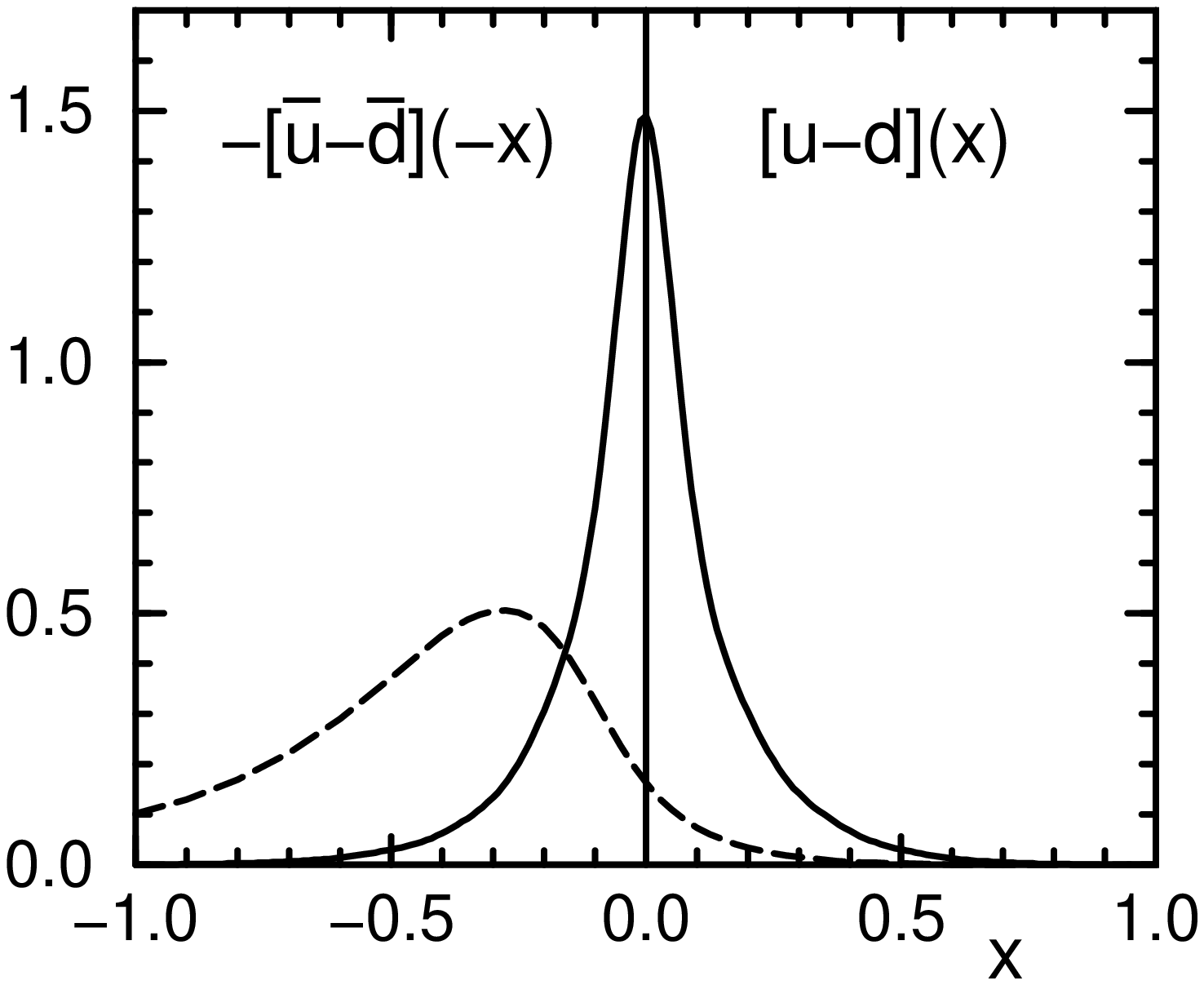}
\caption[]
{The contribution of the Dirac continuum to the isovector distribution
function $u(x) - d(x)$, for $M = 350\,{\rm MeV}$.  At negative $x$ the
function shown describes minus the antiquark distribution. {\it Solid
line:} Total result, as obtained from the expansion in angular velocity,
given by the sum of Eq.(\ref{isovec_combined_1}) and
Eq.(\ref{isovec_combined_2}).  {\it Dashed line:} Result obtained dropping
the contribution Eq.(\ref{isovec_combined_2}), corresponding to the
prescription of Ref.\cite{WK97}.}
\label{fig_fig1}
\end{figure}

The dropping of the contribution Eq.(\ref{isovec_combined_2}) to the parton
distribution function in Ref.\cite{WK97} has rather drastic
consequences. One may expect the differences to the full result to be most
significant at small values of $x$, since they correspond to large
light--like separations of the bilinear operator.  To illustrate the
differences we have performed a numerical calculation of the distribution
function following the prescription of Ref.\cite{WK97} and compared the
results to those of our calculation, where we take into account all terms.
[The details of the numerical calculations and the ultraviolet
regularization are given below in Subsection \ref{subsec_regularization}
and Section \ref{sec_numerical}.]  Fig.\ref{fig_fig1} shows the
contribution of the negative Dirac continuum to the sum over $n$ in in
Eqs.(\ref{isovec_combined_1}, \ref{isovec_combined_2}) in both cases. One
sees that the two distributions differ strongly.  The full result (solid
line), corresponding to the sum of Eqs.(\ref{isovec_combined_1}) and
(\ref{isovec_combined_2}), is large at $x = 0$ and vanishes rapidly for
larger values of $x$. The result obtained after dropping
Eq.(\ref{isovec_combined_2}), as done in Ref.\cite{WK97}, is shown by the
dashed line; in this case the distribution is small at $x = 0$ and does not
vanish for large values of $x$.
\par
To conclude this discussion, we note that the truncations made in
Ref.\cite{WK97} seem problematic in yet another respect. The calculation of
the parton distribution in Ref.\cite{WK97} starts from the bilinear
operator $\bar\psi \ldots \psi$; however, one may just as well take the
opposite ordering of the operator, $-\psi \ldots \bar\psi$.  In QCD the two
orderings give equivalent expressions for the parton distribution, thanks
to the anticommutativity of the fermion fields at space--like
separations. In the chiral quark--soliton model, the two orderings lead,
respectively, to representations of the parton distribution as sums over
occupied and non-occupied quark single--particle states. The equivalence
between these two representations holds only for the full expression for
the distribution as it derives from Eq.(\ref{me_rotation}), {\it i.e.}, for
the sum of {\it all} terms in the expansion with respect to angular
velocity; see Appendix \ref{sec_appendix}.  Consequently, dropping the
contribution Eq.(\ref{isovec_combined_2}) (or otherwise truncating the sums
over levels) one violates this equivalence.  We have verified this in the
numerical calculations. For the full result (the solid line in
Fig.\ref{fig_fig1}) we observe equivalence of summation over positive and
negative energy states, but not for the distribution computed according to
the prescription of Ref.\cite{WK97} (the dashed line in Fig.\ref{fig_fig1})
when computed by summing instead over positive--energy states.
\subsection{Ultraviolet regularization}
\label{subsec_regularization}
In the previous section we have obtained expressions for the isovector
quark distribution as sums over quark single particle levels in the
background pion field. In Eqs.(\ref{isovec_combined_1},
\ref{isovec_combined_2}) the sum over $n$ runs over all occupied levels,
that is, the discrete bound--state level and the negative Dirac continuum,
schematically
\be
\sum\limits_{\scriptstyle n\atop \scriptstyle{\rm occup.}} 
\left[\ldots \right]
&=& \left[ \ldots \right]_{n \; = \; {\rm lev}} + 
\sum\limits_{\scriptstyle n\atop \scriptstyle{\rm neg. cont.}} 
\left[ \ldots \right] ,
\label{level_cont_gen}
\ee
where the ellipsis denotes the ``inner'' sum over $m$ with the restriction
that $E_m \neq E_n$. We shall refer to the two terms on the R.H.S. of
Eq.(\ref{level_cont_gen}) as the ``level'' and ``continuum''
contributions. One should keep in mind, however, that this distinction is
quite formal, and that only the total sum over occupied (or, equivalently,
non-occupied) states has physical significance.
\par
In the derivation of the expressions for the quark distribution function we
have so far not taken into account the ultraviolet cutoff intrinsic in the
effective chiral theory. In fact, the expressions for the isovector
distribution, Eqs.(\ref{isovec_1_occ}, \ref{isovec_2_occ}) {\it viz.}\
Eqs.(\ref{isovec_combined_1}, \ref{isovec_combined_2}), contain an
ultraviolet divergence due to the Dirac continuum contribution, and thus
require regularization. Quite generally, in a calculation of parton
distribution functions there are very strong restrictions on how one should
introduce the UV cutoff in the effective theory. The point is that one has
to preserve certain general properties of parton distributions such as
positivity, sum rules {\it etc.}, which can easily be violated by an
arbitrary UV regularization.  Specifically, the regularization should
preserve the completeness of the set of quark single--particle wave
functions in the soliton pion field. One possible regularization method
which fulfills all requirements is a Pauli--Villars subtraction, which was
used in the calculations of the $N_c$--leading distributions in
Refs.\cite{DPPPW96,DPPPW97,WG97}. We shall also employ this method
here. Thus, we regularize Eqs.(\ref{isovec_1_occ}, \ref{isovec_2_occ}) {\it
viz.}\ Eqs.(\ref{isovec_combined_1}, \ref{isovec_combined_2}) as follows:
\be
[u(x) - d(x)]_{\rm reg.}
&=&  
\left[ \ldots \right]_{n \; = \; {\rm lev}, M} + 
\left(
\sum\limits_{\scriptstyle n\atop \scriptstyle{\rm neg. cont.}} 
\left[\ldots \right]_M
- \frac{M^2}{M_{PV}^2}
\sum\limits_{\scriptstyle n\atop \scriptstyle{\rm neg. cont.}} 
\left[\ldots \right]_{M_{PV}}
\right) ,
\label{PV} 
\ee
where the subscript $M_{PV}$ denotes the corresponding expression computed
with the constituent quark mass, $M$, replaced by the regulator mass,
$M_{PV}$, {\it cf.}\ Eqs.(\ref{E_tot_reg}, \ref{fpi}).  This subtraction
removes the logarithmic divergence of the distribution function, as can be
shown using methods similar to those developed in Ref.\cite{DPPPW97}
(gradient expansion).  We do not include the contribution of the discrete
level to the isovector distribution in the Pauli--Villars subtraction; this
contribution is finite and does not need to be subtracted.
\par
The moment of inertia of the soliton, Eq.(\ref{I}), is also ultraviolet
divergent. It must be regularized consistently with the isovector
distribution in order to preserve the isospin sum rule (see Subsection
\ref{subsec_isospin}). This is achieved by regularizing it by a
Pauli--Villars subtraction analogous to Eq.(\ref{PV}).
\par
We stress again that the calculation of parton distributions in the
effective chiral theory, {\it i.e.}, the identification of the twist--2 QCD
operator with an operator expressed in terms of fields of the effective
theory, Eq.(\ref{nonlocal}), is based on the parametric smallness of the
ratio of the constituent quark mass, $M$, to the ultraviolet cutoff (here:
$M_{PV}$). The approach is consistent in the sense that the expressions for
the parton distributions in the effective theory are logarithmically
divergent and thus insensitive to the details of the ultraviolet
regularization (assuming that the latter does not violate any important
properties such as completeness {\it etc.}). It should be noted, however,
that the lack of precise knowledge of the ``true'' ultraviolet
regularization leads to a theoretical uncertainty at the level of finite
terms, $\propto M^2 / M_{PV}^2$. For instance, Pauli--Villars subtraction
of the finite level contribution would correspond to a modification of the
distribution function by an amount of order $M^2 / M_{PV}^2$.  The
numerical results presented in Section \ref{sec_numerical} should be
understood as a calculation of the parton distributions with a typical
regularization fulfilling all general requirements, with the theoretical
uncertainty in the ultraviolet regularization leading to a numerical
uncertainty of typically 10--20 \% --- except for the sum rules, which
follow from general principles not violated by the ultraviolet
regularization.
\subsection{The isospin sum rule}
\label{subsec_isospin}
We now want to demonstrate that the expressions derived for the isovector
distribution in the chiral quark--soliton model are consistent with the
isospin sum rule,
\be
\int\limits_{-1}^{1} dx [u(x) - d(x)]
&\equiv& \int\limits_{0}^{1} dx [u(x) - \bar u(x) -  d(x) + \bar d(x)]
=  2 T_3 .
\label{isospin_sumrule}
\ee
Consider the representation of the isovector distribution function as a sum
over quark levels, Eqs.(\ref{isovec_combined_1}, \ref{isovec_combined_2}).
Since the contribution $[u(x)-d(x)]^{(2')}$, Eq.(\ref{isovec_combined_2}),
is a total derivative, it does not contribute to the integral in
Eq.(\ref{isospin_sumrule}), and we can concentrate on the contribution
$[u(x)-d(x)]^{(1')}$, Eq.(\ref{isovec_combined_1}).  Integrating
Eq.(\ref{isovec_combined_1}) over $x$, replacing in the large--$N_c$ limit
the integral from $-1$ to $1$ by the integral over the whole real axis, we
use up the delta functions in Eq.(\ref{isovec_combined_2}) and obtain
\be
\int\limits_{-1}^{1} dx
[u(x) - d(x)]
&=& (2T^3) \frac{N_c}{6 I} 
\sum\limits_{\scriptstyle n\atop \scriptstyle{\rm occup.}}
\sum\limits_{\scriptstyle m\atop {\scriptstyle {E_m \neq E_n}}}
\frac{\langle n|\tau ^a|m\rangle
\langle m|\tau ^a |n \rangle}{E_m - E_n} .
\ee
The hedgehog symmetry of the pion field, Eq.(\ref{hedge}), has allowed us
to drop the $\gamma^0\gamma^3$--term. Taking into account that the soliton
moment of inertia, Eq.(\ref{I}), can equivalently be written as
\be
I &=& \frac{N_c}{6} 
\sum\limits_{\scriptstyle n \atop \scriptstyle {\rm occup.}}
\sum\limits_{\scriptstyle m\atop {\scriptstyle {E_m \neq E_n}}}
\frac{\langle n | \tau^a | m \rangle \langle m | \tau^a | n \rangle}
{E_m - E_n} ,
\label{I_all}
\ee
we immediately reproduce the isospin sum rule,
Eq.(\ref{isospin_sumrule}). Also, it can be seen that the
Pauli--Villars regularization of the distribution function and the moment
of inertia as defined by Eq.(\ref{PV}) does not upset this proof; in
particular, it preserves the equivalence of the two representations of the
moment of inertia, Eq.(\ref{I}) and Eq.(\ref{I_all}).
\subsection{The Gottfried sum}
\label{subsec_gottfried}
An interesting quantity related to the isovector antiquark distribution is
the Gottfried sum \cite{Gottfried67,Kumano97}, which is defined as
\be
I_G &=& \frac{1}{3} + \frac{2}{3} \int_0^1 dx \ [\bar{u}(x) - \bar{d}(x)] .
\label{Gottfried_sum}
\ee
The integral on the R.H.S.\ is scale--dependent only at two--loop level;
its scale dependence is negligible over the entire perturbative region. If
the sea quark distribution were isospin symmetric, 
$\bar{u}(x) = \bar{d}(x)$, which would be the case if, for example, one
assumed the sea quark distribution to be generated entirely radiatively,
this quantity would be equal to $1/3$ (Gottfried sum rule).  However, the
NMC experiment \cite{NMC94} finds a significant deviation from this value,
\be
I_G &=& 0.235 \pm 0.026 
\hspace{1cm} \mbox{at} \hspace{1cm} Q^2 \;\; = \;\; 4\,{\rm GeV}^2 ,
\ee
indicating that the sea quark distribution is rather far from
flavor--symmetric. Note that the Gottfried sum rule does not follow from
any fundamental principles of QCD. In fact, the large--$N_c$ picture of the
nucleon as a chiral soliton naturally explains the presence of a
flavor--nonsymmetric antiquark distribution.
\par
The expression for the Gottfried sum in the chiral quark--soliton model is
obtained by integrating the expression for the total isovector quark
distribution, Eqs.(\ref{isovec_combined_1}, \ref{isovec_combined_2}), over
$x$ from $-1$ to $0$, keeping in mind that the antiquark distribution is
given by minus the expressions in Eqs.(\ref{isovec_combined_1},
\ref{isovec_combined_2}) at $-x$. Since the integration extends only over
half of the $x$--axis, also the total derivative term,
Eq.(\ref{isovec_combined_2}), contributes to the Gottfried sum, in contrast
to the isospin sum rule, Eq.(\ref{isospin_sumrule}), where the
integration runs over the whole $x$--axis and Eq.(\ref{isovec_2_occ}) drops
out. This contribution to the Gottfried sum in the chiral quark--soliton
model has not been mentioned in Ref.\cite{DPPPW96}.
\par
We refrain from writing down the analytic expressions for the Gottfried sum
obtained by integrating Eqs.(\ref{isovec_1_occ},
\ref{isovec_2_occ}). Instead, we shall compute Eq.(\ref{Gottfried_sum}) by
integration of the numerically computed antiquark distribution function,
see below. In particular, the ultraviolet regularization of the Gottfried
sum and all other properties will follow directly from that of the
distribution functions.
\par
We note that the Gottfried sum in the large--$N_c$ limit has been discussed
previously in the context of the chiral quark-soliton model
\cite{Wakamatsu92} and the Skyrme model \cite{WH93}. These studies were,
however, based on expressions for the R.H.S.\ of Eq.(\ref{Gottfried_sum})
which do not follow from a consistent identification of the isovector
distribution function within the effective model.
\section{Numerical results and discussion}
\setcounter{equation}{0}
\label{sec_numerical}
The method we choose for the numerical computation of the isovector
unpolarized distribution function parallels that for computing the
isosinglet unpolarized one \cite{DPPPW97}.  We evaluate the sums over quark
single--particle levels in the background pion field by placing the soliton
in a spherical 3--dimensional box, where the eigenfunctions of the Dirac
Hamiltonian in the background pion field can be obtained by numerical
diagonalization \cite{KR84}. To facilitate the calculations we first
convert the expressions Eqs.(\ref{isovec_combined_1},
\ref{isovec_combined_2}) to a spherically symmetric form by averaging over
the 3--dimensional orientations of the spatial separation implied in the
non-local operator Eq.(\ref{nonlocal}), see Ref.\cite{DPPPW97} for
details. In the spherically symmetric version of
Eqs.(\ref{isovec_combined_1}, \ref{isovec_combined_2}) the matrix elements
of the operators between single--particle quark states can easily be
computed using the standard angular--momentum selection rules.  The
discontinuous functions of the single--particle energies and the
single--particle momentum operator, which arise as a consequence of the
presence in Eqs.(\ref{isovec_combined_1}, \ref{isovec_combined_2}) of the
functions $\delta (E_n + P^3 - x M_N)$, are smoothed using the ``smearing''
method of Ref.\cite{DPPPW97}.\footnote{The contribution
Eq.(\ref{isovec_combined_1}) to the isovector distribution function is a
double sum over quark single--particle levels, similar to the moment of
inertia, Eq.(\ref{I}). In the evaluation of these quantities in a finite
box there arises the complication that the boundary conditions which have
to be imposed on the single--particle wave functions can lead to spurious
``vacuum'' contributions. In Ref.\cite{WY91} a method has been devised to
circumvent this problem by using two sets of basis functions in the box
subject to different boundary conditions. We have employed this technique
in the calculations of distribution functions reported here.}
\par
For the ultraviolet regularization of distribution functions we employ the
Pauli--Villars subtraction, Eq.(\ref{PV}). For the sake of numerical
stability we first evaluate the sums over levels
Eqs.(\ref{isovec_combined_1}, \ref{isovec_combined_2}) with a smooth energy
cutoff, both the sums over levels obtained with the usual quark mass, $M$,
and the regulator mass, $M_{PV}$.  We then perform the Pauli--Villars
subtraction, Eq.(\ref{PV}), and remove the energy cutoff by extrapolation
to infinity.
\par
\begin{figure}
\setlength{\epsfxsize}{15cm}
\setlength{\epsfysize}{15cm}
\epsffile{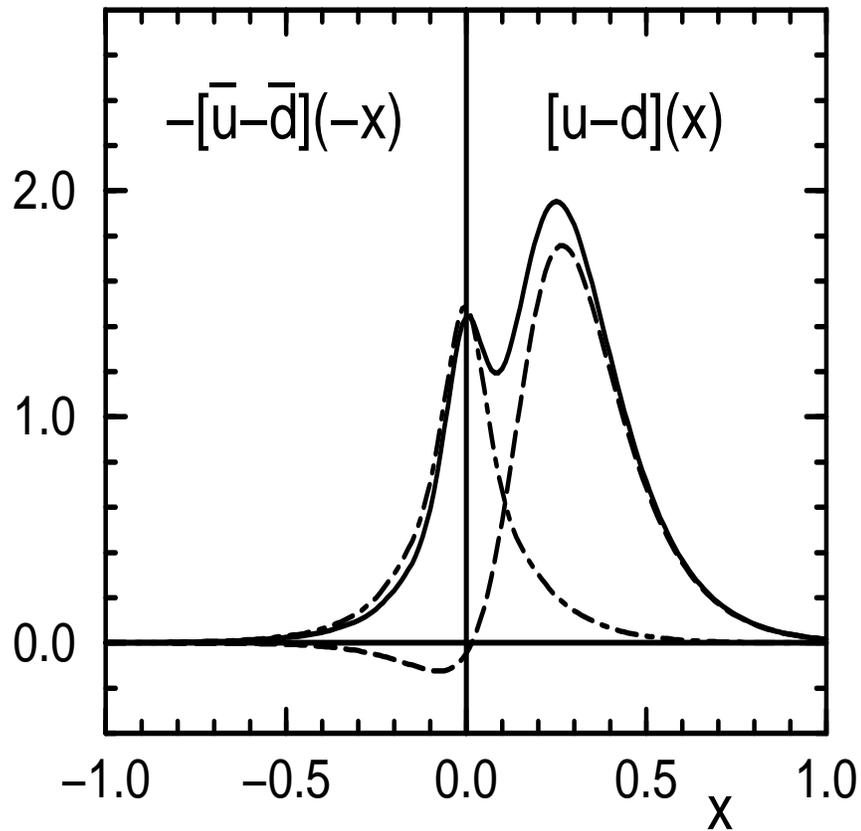}
\caption[]
{Contributions to the isovector unpolarized distribution function, 
$u(x) - d(x)$, for $M = 350\,{\rm MeV}$. At negative $x$ the function shown
describes minus the antiquark distribution.  {\it Dashed line:}
Contribution of the discrete level.  {\it Dot--dashed line:} Contribution
of the negative Dirac continuum.  {\it Solid line:} Total result.}
\label{fig_fig2}
\end{figure}
\begin{figure}
\setlength{\epsfxsize}{15cm}
\setlength{\epsfysize}{15cm}
\epsffile{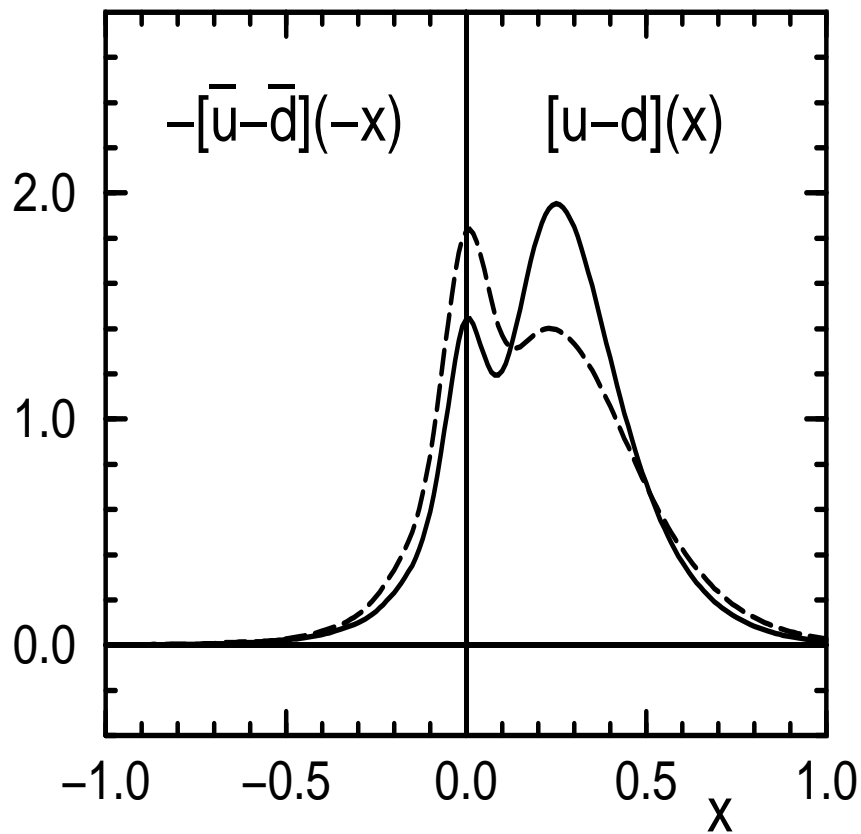}
\caption[]
{The isovector unpolarized distribution function, $u(x) - d(x)$, for
constituent quark masses $M = 350\,{\rm MeV}$ ({\it solid line}) and 
$M = 420\,{\rm MeV}$ ({\it dashed line}).  At negative $x$ the function
shown describes minus the antiquark distribution.}
\label{fig_fig3}
\end{figure}
\begin{figure}
\setlength{\epsfxsize}{15cm}
\setlength{\epsfysize}{15cm}
\epsffile{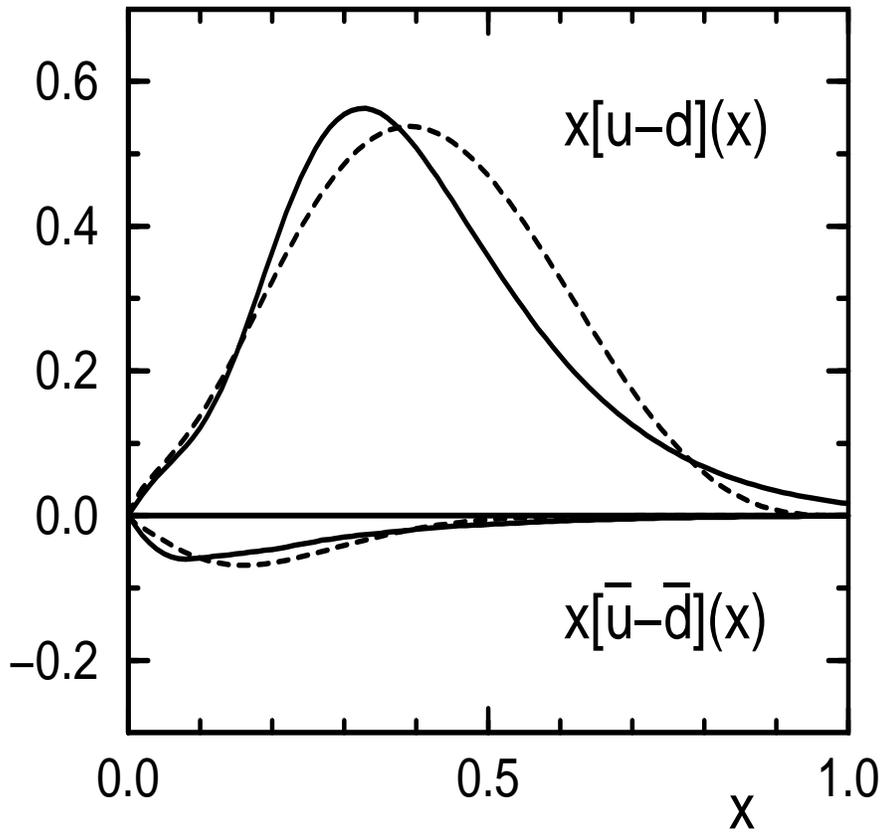}
\caption[]
{{\it Solid lines}: The calculated isovector unpolarized quark-- and
antiquark distributions, $x[u(x) - d(x)]$ and $x[\bar u(x) - \bar d(x)]$,
for $M = 350\,{\rm MeV}$. Shown is the total result, corresponding to the
solid line in Fig.\ref{fig_fig2}. {\it Dotted lines:} The GRV NLO
parametrizations \cite{GRV95}.}
\label{fig_fig4}
\end{figure}

We have calculated the isovector quark distribution for two values of the
constituent quark mass, $M = 350\, {\rm MeV}$, which is the value obtained
in Ref.\cite{DP86} for the instanton vacuum, and $M = 420\, {\rm MeV}$. The
ultraviolet cutoff is in both cases determined by fitting the pion decay
constant, Eq.(\ref{fpi}), $M_{PV}^2/M^2 = 2.52\; (M = 350\,{\rm MeV})$ and
$M_{PV}^2/M^2 = 1.90\; (M = 420\,{\rm MeV})$.  The soliton profile,
Eq.(\ref{hedge}), and the nucleon mass have been found by self--consistent
minimization of the Pauli--Villars regularized static energy in
Ref.\cite{WG97}, $M_N = 1140\,{\rm MeV}\; (M = 350\,{\rm MeV})$ and 
$M_N = 1040\,{\rm MeV}\; (M = 420\,{\rm MeV})$.  A number of other hadronic
nucleon observables such as the isovector axial coupling constant
$g_A^{(3)}$ have been calculated with this ultraviolet regularization in
Ref.\cite{Doering92}.  The results for the isovector unpolarized quark and
antiquark distributions are shown in Figs.\ref{fig_fig2}--\ref{fig_fig4}.
\par
Before comparing our results with the parametrizations of the experimental
data it is instructive to study the behavior of the different contributions
to the distribution function in this model. Fig.\ref{fig_fig2} shows the
function $u(x) - d(x)$ for both positive and negative $x$, describing the
isovector quark distribution at positive and minus the antiquark
distribution at negative values of $x$. The dashed line shows the
contribution of the discrete level, as defined by
Eq.(\ref{level_cont_gen}). This contribution is concentrated around values
$x \sim 1/3$ and similar in shape to the bound--state level contribution to
the isosinglet unpolarized and isovector polarized distributions
\cite{DPPPW97,WG97}.  The contribution of the Dirac continuum, {\it cf.}\
Eq.(\ref{level_cont_gen}), is shown by the dot--dashed line. Similar to the
isosinglet distribution this contribution is peaked around $x = 0$;
however, in the isovector case this function does not change sign at 
$x = 0$.  The total is given by the solid line; one observes that it is
essentially a smooth function except for a region of $x$ around $x = 0$,
where it is dominated by the Dirac continuum contribution.  Note also that
the calculated distribution satisfies the isospin sum rule, as discussed in
Subsection \ref{subsec_isospin} (the area under the solid line in
Fig.\ref{fig_fig2} is unity).
\par
The isovector quark and antiquark distributions calculated here refer to a
low normalization point of the order of the cutoff of the effective chiral
theory ($\simeq 600\,{\rm MeV}$), and can be compared to experimental data
for structure functions only after perturbative evolution to larger
scales. Hence the ``small--$x$'' behavior of the calculated parton
distribution apparent from Fig.\ref{fig_fig2} has significance only in the
sense of an input distribution at a low scale and does not imply a
statement about the small--$x$ behavior of the structure functions at
experimental scales.
\par
The parton distribution functions calculated in the effective theory are
but logarithmically divergent with the UV cutoff and thus, for typical
values of $x$, insensitive to the details of the UV cutoff (assuming a
physically acceptable regularization scheme meeting the criteria discussed
above). However, it is known that in the parametrically small region 
$|x| \le (M /\Lambda )^2 / N_c$ the shape of the distribution depends on
the details of the UV regularization, as has recently been discussed in
connection with the calculation of off--forward parton distributions in
this approach \cite{PPPBGW97}. For instance, it was seen there that the
discontinuity at $x = 0$ of the isosinglet distribution calculated with
Pauli--Villars regularization is reduced to a smooth transition when the
regularization is implemented in the form of a momentum--dependent
constituent quark mass, as suggested by the instanton vacuum.  [For values
of $x$ not in the vicinity of zero results are practically unchanged as
compared to the Pauli--Villars regularization.]  One may thus surmise that
also the shape of the peak in the isovector distribution at $x = 0$ depends
strongly on the ultraviolet cutoff. The calculation of the
$1/N_c$--suppressed distribution functions with regularization by a
momentum--dependent mass is outside the scope of the presently available
methods.  The interesting general problem of the relation of the
ultraviolet cutoff of the effective theory to the ``small--$x$'' behavior
of the low--scale parton distributions will be left for further
investigation.
\par
In Fig.\ref{fig_fig3} we compare the distribution functions obtained for
two different values of the constituent quark mass, $M = 350 \, {\rm MeV}$
and $M = 420 \, {\rm MeV}$. With the ultraviolet regularization chosen
according to Eqs.(\ref{E_tot_reg}, \ref{fpi}) and Eq.(\ref{PV}), and the
soliton profile determined by minimization of the energy
Eq.(\ref{E_tot_reg}), the only remaining free parameter in this model
calculation is the value of the constituent quark mass, $M$. Rather than
trying to determine an ``optimum'' value by performing a best fit to a
number of hadronic observables, it is more interesting here to study the
dependence of the calculated parton distributions on this parameter.
Fig.\ref{fig_fig3} gives an idea of the variation of the results with $M$.
\par
The parton distributions computed here should be used as input for
perturbative evolution, starting with a scale of the order of the cutoff
$(\simeq \, {\rm 600 MeV})$. We shall not perform the evolution here, but
rather compare our results with the parameterizations by Gl\"{u}ck, Reya,
and Vogt \cite{GRV95}. These authors generate parton distributions at
experimental $Q^2$ from ``valence--like'' (non-singular) input quark--,
antiquark and gluon distributions at a normalization point of the order of
$600 \, {\rm MeV}$ and obtain excellent fits to the data from
deep--inelastic scattering and a variety of other processes. In
Fig.\ref{fig_fig4} we compare the isovector distribution of quarks, 
$x[u(x) - d(x)]$, and antiquarks, $x[\bar u(x) - \bar d(x)]$, to the GRV
parametrization \cite{GRV95}.  One notes that the calculated distributions
are systematically harder ({\it i.e.}, centered at larger values of $x$)
than the GRV one, indicating that the normalization point of the calculated
distributions is even lower than that of the GRV parametrization. This is
consistent with the conclusions drawn from the comparison of the isosinglet
unpolarized \cite{DPPPW96,DPPPW97,WG97} and isovector polarized
\cite{DPPPW96,DPPPW97} distributions to the GRV parametrizations; however,
in the isovector unpolarized case considered here the comparison of the
calculated distribution with the GRV fit is more direct since, contrary to
the isosinglet unpolarized distribution, this distribution does not mix
with the gluon distribution under evolution, and its normalization is
scale--independent. [Also, contrary to the isovector polarized distribution
whose normalization is given by the isovector axial coupling constant,
$g_A^{(3)}$, whose value in our approach is model--dependent, the
normalization of the the isovector unpolarized distribution is universal
thanks to the isospin sum rule.]
\par
Note that the order of magnitude and shape of the calculated isovector
antiquark distribution are in good agreement with the GRV parametrization,
see Fig.\ref{fig_fig4}. As can be seen from Fig.\ref{fig_fig2} the
antiquark distribution is dominated by the Dirac continuum contribution. We
note that our distribution differs significantly from the result of
Ref.\cite{WK97}, where the contribution from the term
Eq.(\ref{isovec_combined_2}) has been dropped; our distribution vanishes
rapidly for large $x$, in contrast to the one of Ref.\cite{WK97}, see
Fig.\ref{fig_fig1}.
\begin{table}
\begin{center}
\begin{tabular}{|c|c|c|c||c|} \hline
$M$ & \multicolumn{3}{c||}{$\int_0^1 dx \ [\bar u (x) - \bar d (x)]$} 
& $I_G$ 
\\[.3cm]
    & level & continuum & total & \\ \hline
$350\, {\rm MeV}$ &  0.030 & -0.201 & -0.171  &  0.219  \\ \hline
$420\, {\rm MeV}$ &  0.049 & -0.281 & -0.232  &  0.178  \\ \hline
\end{tabular}
\end{center}
\caption{The integral of the antiquark distribution and the 
Gottfried sum, Eq.(\ref{Gottfried_sum}),
for constituent quark masses $350\, {\rm MeV}$ and $420\, {\rm MeV}$.
{\it Columns 2--4:} integral of the antiquark distribution, 
$\int_0^1 dx \ [\bar{u}(x) - \bar{d}(x)]$: contribution of the
discrete level, the Dirac continuum and the total result,
{\it cf.}\ Fig.\ref{fig_fig2}. {\it Column 5:} Gottfried sum,
Eq.(\ref{Gottfried_sum}). The experimental value quoted in 
Ref.\cite{NMC94} is $I_G = 0.235 \pm 0.026$ at 
$Q^2 = 4\,{\rm GeV}^2$.}
\label{table_1}
\end{table}
\par
With regard to the isovector antiquark distribution it is interesting to
compute the Gottfried sum, Eq.(\ref{Gottfried_sum}). The results for the
two constituent quark masses $M = 350\, {\rm MeV}$ and 
$M = 420\, {\rm MeV}$ are given in Table \ref{table_1}, where we also list
separately the contributions of the discrete level and the Dirac continuum
to the integral from of the antiquark distribution. As can be seen, the
deviation of the Gottfried sum from from $1/3$ is dominated by the Dirac
continuum contribution. The values are in reasonable agreement with the NMC
value \cite{NMC94}. We note that, since our Dirac continuum contribution
strongly differs from that of Ref.\cite{WK97}, the good agreement of the
value of $I_G$ reported there with experiment seems fortuitous.
\par
Finally, we note that the parton distributions studied here refer to the
large--$N_c$ limit, where the nucleon is heavy, $M_N \propto N_c$. The
calculated distributions therefore do not go to zero at $x = 1$, rather,
they are exponentially small at large $x$, as discussed in
Ref.\cite{DPPPW96}.  For this reason we have evaluated the integral of the
antiquark distribution in Eq.(\ref{Gottfried_sum}) and Table \ref{table_1}
with the upper limit replaced by infinity. Numerically, the contribution
from values of $x > 1$ to the integral is negligible.
\section{Conclusions}
\setcounter{equation}{0}
\label{sec_conclusions}
In this paper we have extended the approach to calculate parton
distributions at a low normalization point in the large--$N_c$ limit to the
case of the distributions which appear in the subleading order of the
$1/N_c$--expansion. Specifically, we have computed the isovector
unpolarized distribution of quarks and antiquarks.  The methods developed
here can readily be generalized to compute also the other
$1/N_c$--subleading distributions, namely the isosinglet longitudinally and
transverse polarized ones, as well as the rotational $1/N_c$--corrections
to the isovector polarized distribution.  These distributions are currently
being computed.
\par
We have found reasonable agreement of the calculated distributions with the
parametrizations of the data at a low normalization point.  In particular,
the large--$N_c$ approach naturally predicts a flavor--asymmetric ({\it
i.e.}, non-radiative) sea quark distribution of correct sign and
magnitude. The $x$--dependence of the isovector antiquark distribution
compares well with the GRV parametrizations. It is interesting that the
calculated distributions exhibit a strong rise at $x = 0$.  Also, the
integral of the calculated isovector antiquark distribution (the violation
of the Gottfried sum rule) is compatible with the experimental data, in
view of the general theoretical uncertainties of the present approach.
\\[1.2cm]
{\large\bf Acknowledgements}
\\[.3cm]
The authors are deeply grateful to D.I.\ Diakonov and V.Yu.\ Petrov for
many enlightning conversations. 
\\[.2cm]
This work has been supported in part by a
joint grant of the Russian Foundation for Basic Research (RFBR) and the
Deutsche Forschungsgemeinschaft (DFG) 436 RUS 113/181/0 (R), by
RFBR grant 96-15-96764, by the NATO Scientific Exchange grant
OIUR.LG 951035, by INTAS grants 93-0283 EXT and 93-1630-EXT, 
by the DFG and by COSY (J\"ulich). The Russian participants acknowledge 
the hospitality of Bochum University.
%
%
\appendix
\renewcommand{\theequation}{\Alph{section}.\arabic{equation}}
\setcounter{equation}{0}
\section{Evaluation of the isovector distribution}
\setcounter{equation}{0}
\label{sec_appendix}
In this appendix we evaluate the two contributions to the isovector
distribution arising from the expansion in angular frequency,
Eqs.(\ref{omegaexp_1}) and (\ref{omegaexp_2}).  Our aim is to express these
contributions in the form of sums over quark single--particle levels, which
{\it e.g.}\ can serve as a starting point for a numerical calculation of
the distribution function.
\par
We begin by rewriting the contributions Eqs.(\ref{omegaexp_1}) and
(\ref{omegaexp_2}) in the form
\be
\lefteqn{[u(x)-d(x)]^{(1)} \;\; = \;\; } && \nonumber \\
&& - \frac{N_cM_N i}{8\pi I} \sum\limits_{S_3}
\int\limits_{-\infty}^\infty dz^0  e^{ixM_Nz^0} \int d^3{\bf X} 
\int dR \, \left[\phi_{T_3S_3}^{T=S}(R)\right]^\ast
D_{3b}(R) S^a \phi_{T_3S_3}^{T=S}(R)
\nonumber \\
&&\times \; \mbox{Tr} \left[  \tau^b (1+\gamma^0\gamma^3)
\langle {z^0,{\bf z}-{\bf X}} |
\frac1{i\partial_t - H(U_c)}\, \tau^a
\frac1{i\partial_t - H(U_c)}
|{0,-{\bf X}} \rangle  \right]
\Bigr|_{z^3=-z^0,\>z_\perp=0} ,
\nonumber \\
\ee
\be
\lefteqn{[u(x)-d(x)]^{(2)} \;\; = \;\;} && \nonumber \\
&& \frac{M_N N_c}{8\pi I} \sum\limits_{S_3}
\int\limits_{-\infty}^\infty dz^0  e^{ixM_Nz^0}
\int d^3{\bf X} \int dR \,
\left[\phi_{T_3S_3}^{T=S}(R)\right]^\ast
z^0 D_{3b}(R) S^c \nonumber \\
&& \times \; \mbox{Tr} \left[ \tau^b\tau^c
(1+\gamma^0\gamma^3)
\langle {z^0,{\bf z}-{\bf X}} |
\frac1{i\partial_t - H(U_c)}
|{0,-{\bf X}} \rangle \right]
\Bigr|_{z^3=-z^0,\>z_\perp=0}
\phi_{T_3S_3}^{T=S}(R) ,
\ee
where $\mbox{Tr}\ldots$ denotes the trace over Dirac and flavor indices.
Here we have introduced the Wigner $D$--function in the vector
representation
\be
D_{ab}(R) &=& \frac{1}{2} \mbox{Tr} 
\left( \tau_a R \tau_b R^\dagger \right) .
\ee
We first have to compute the rotational matrix element. Strictly speaking,
this matrix element contains noncommuting operators, $D_{3b}(R)$ and $S^a$,
and one should be careful about their ordering. However, due to the average
over the nucleon spin the result does not depend on the order, and one has
\be
&&
\sum\limits_{S_3}
\int dR \,
\left[\phi_{T_3S_3}^{T=S}(R)\right]^\ast
D_{3b}(R) S^a \phi_{T_3S_3}^{T=S}(R)
\nonumber\\
&=&
\sum\limits_{S_3}
\int dR \,
\left[\phi_{T_3S_3}^{T=S}(R)\right]^\ast
S^a D_{3b}(R)  \phi_{T_3S_3}^{T=S}(R)
\;\; = \;\; - \frac{1}{3} \delta^{ab} (2T^3) .
\ee
Using this, and passing from the time to the frequency representation for
the quark Green functions we arrive at
\be
\lefteqn{
[u(x)-d(x)]^{(1)} \;\; = \;\;
(2T^3) \frac{i N_cM_N}{24\pi I} 
\int\limits_{-\infty}^\infty dz^0  e^{ixM_Nz^0}
\int \frac{d\omega}{2\pi} e^{-i\omega z^0} 
\int d^3{\bf X} }
&& \nonumber \\
&& \times \; \mbox{Tr} 
\left[ \tau^a (1 + \gamma^0\gamma^3)
\langle{{\bf z}-{\bf X}} |
\frac1{\omega - H(U_c)}\, \tau^a
\frac1{\omega - H(U_c)}
|{-{\bf X}} \rangle  \right]
\Bigr|_{z^3=-z^0,\>z_\perp=0} ,
\ee
\be
\lefteqn{[u(x)-d(x)]^{(2)} \;\; = \;\;
- (2T^3) \frac{N_c M_N}{8\pi I} 
\int\limits_{-\infty}^\infty dz^0  e^{ixM_Nz^0} z^0 
\int \frac{d\omega}{2\pi} e^{-i\omega z^0} \int d^3{\bf X} }
&& \nonumber \\
&& \times
\; \mbox{Tr} \left[
(1 + \gamma^0\gamma^3)
\langle{{\bf z}-{\bf X}} |
\frac1{\omega - H(U_c)}
|{-{\bf X}} \rangle \right]
\Bigr|_{z^3=-z^0,\>z_\perp=0} .
\ee
[The treatment of the poles in the $\omega$--integral implied here will be
described below.] These expressions have the form of an integral over all
space of a matrix element of the quark propagator between position
eigenstates localized at different points. It is convenient to rewrite them
as functional traces ({\it i.e.}, sums over diagonal matrix elements) by
introducing the finite translation operator, which is given as the
exponential of the single--particle momentum operator, $P^k$,
\be
\langle{{\bf z}-{\bf X}} |
&=& \langle{-{\bf X}}| \exp (i P^k z^k ) .
\ee
We can then perform the integral over $z^0$ and write the result in the
form
\be
\lefteqn{[u(x)-d(x)]^{(1)} \;\; = \;\;} && \nonumber \\[.5em]
&& (2T^3) \frac{i N_cM_N}{24\pi I} \int\limits_{-\infty}^{\infty} d\omega
\; \mbox{Sp} \left[ \tau^a
(1+\gamma^0\gamma^3)
\delta(\omega+P^3 - xM_N)
\frac1{\omega - H(U_c)}\, \tau^a
\frac1{\omega - H(U_c)} 
\right] ,
\nonumber \\
\label{isovec_1_trace}
\ee
\be
[u(x)-d(x)]^{(2)} &=& (2T^3) 
\frac{i N_c}{8\pi I} \frac{\partial}{\partial x}
\int\limits_{-\infty}^{\infty} d\omega \;
\mbox{Sp} \left[ (1 + \gamma^0\gamma^3)   
\delta(\omega+P^3 - xM_N) \frac1{\omega - H(U_c)} \right] ,
\nonumber \\
\label{isovec_2_trace}
\ee
where $\mbox{Sp}\ldots$ denotes the functional trace in the space of
single--particle quark states. These functional traces can now be computed
using a basis of eigenstates of the Dirac Hamiltonian in the background
pion field, Eq.(\ref{eigen_n}). As to the second contribution,
Eq.(\ref{isovec_2_trace}), one may easily show that, up to a factor, it is
simply the derivative in $x$ of the isosinglet distribution function, which
appears in the leading order of the $1/N_c$--expansion
\cite{DPPPW96,DPPPW97}. It can be written as a simple sum over occupied
single--particle levels:
\be
[u(x)-d(x)]^{(2)}
&=& - (2T^3) \frac{1}{4IM_N} \times \frac{\partial}{\partial x}
[u(x)+d(x)]^{\rm leading} \nonumber \\
&=& - (2T^3) \frac{ N_c}{4I} \times \frac{\partial}{\partial x}
\sum_{\scriptstyle n \atop \scriptstyle {\rm occup.}}
\langle{n}| (1+\gamma^0\gamma^3)   
\delta(E_n + P^3 - xM_N) |{n} \rangle .
\nonumber \\
\ee
It is also possible to express this quantity as a sum over non-occupied
levels, Eq.(\ref{isovec_2_nonocc}), see Refs.\cite{DPPPW96,DPPPW97}.
\par
Let us now consider the first contribution, Eq.(\ref{isovec_1_trace}). It
gives rise to a double sum over levels,
\be
[u(x)-d(x)]^{(1)} &=& (2T_3 )
\frac{i N_cM_N}{24\pi I} \int\limits_{-\infty}^{\infty} d\omega
\sum\limits_{m,n}
\langle n | \tau^a(1+\gamma^0\gamma^3) \delta(\omega+P^3 - xM_N) | m\rangle
\nonumber \\
&& \times \langle m | \tau^a |n \rangle
\frac1{(\omega - E_m)(\omega - E_n)} ,
\label{isovec_1_doublesum}
\ee
where the sums over $n, m$ run over all levels.  We can further simplify
the form of Eq.(\ref{isovec_1_doublesum}) by taking into account that the
expression on the R.H.S.\ is real.  First, this is natural, since the quark
distribution function, $u(x)-d(x)$, must be real. More formally, one may
prove the reality of the R.H.S.\ of Eq.(\ref{isovec_1_doublesum}) starting
from the locality of the Dirac Hamiltonian in the background field, and the
fact that neither the angular velocity corrections to the Hamiltonian nor
our ultraviolet regularization violate this property. We may thus replace
the expression on the R.H.S.\ of Eq.(\ref{isovec_1_doublesum}) by its real
part:
\be
\lefteqn{ [u(x)-d(x)]^{(1)} \;\; = \;\;} && 
\nonumber \\
&& -(2T^3)\frac{N_cM_N}{24\pi I} \; \mbox{Im} \;
\int\limits_{-\infty }^\infty d\omega
\sum\limits_{m,n}\langle n|\tau ^a(1+\gamma
^0\gamma ^3)\delta (\omega +P^3-xM_N)|m\rangle
\nonumber \\
&& \times \langle m|\tau ^a|n\rangle \frac 1{\left[ \omega -E_m+i0\eta
(E_m)\right] \left[ \omega -E_n+i0\eta (E_n)\right] } .
\label{isovec_1_real}
\ee
Here the poles in the $\omega$--integral corresponding to the
single--particle energies are shifted according to the occupation of the
levels
\be
\eta ( E_n ) &=& \mbox{sign} (E_n - E_{\rm lev} - 0) .
\label{eta_def}
\ee
At $E_m \neq E_n$ we can write
\be
\lefteqn{
\mbox{Im}\;\frac 1{\left[ \omega -E_m+i0\eta (E_m)\right] \left[ \omega
-E_n+i0\eta (E_n)\right] } } &&
\nonumber \\
&=& -\pi \eta (E_m)\delta (\omega -E_m)\frac 1{\omega -E_n} \; - \; 
\pi \eta (E_m)\delta (\omega -E_m)\frac 1{\omega -E_n}
\nonumber \\
&=& 
-\pi [\eta (E_m)\delta (\omega -E_m) \; - \; 
\eta (E_n)\delta (\omega -E_n)]
\frac{1}{E_m-E_n} . 
\label{impart_diff}
\ee
What to do in the case $E_m = E_n$ depends on whether we consider the case
of infinite volume, where most of the spectrum is continuous, or whether we
work in a large but finite box, in which case the spectrum is
quasi--discrete. We concentrate on the latter case, since it is relevant
for the numerical calculations.  In the quasi--discrete case, if 
$E_m = E_n$ one should write instead of Eq.(\ref{impart_diff}):
\be
\lefteqn{
\left. \mbox{Im}\;
\frac 1{\left[ \omega -E_m+i0\eta (E_m)\right] \left[ \omega
-E_n+i0\eta (E_n)\right] }\right| _{E_m=E_n} } && \nonumber \\
&=&
\mbox{Im}\;
\frac 1{\left[ \omega -E_m+i0\eta (E_m)\right] ^2} \;\;= \;\;
-\frac \partial
{\partial \omega }\mbox{Im}\frac 1{\omega -E_m+i0\eta (E_m)}
\nonumber \\
&=& 
\frac \partial {\partial \omega }\pi \eta (E_m)\delta (\omega -E_m)
\;\; = \;\;
\pi \eta (E_m)\delta ^{\prime }(\omega -E_m)  
\label{impart_same} .
\ee
Note that the same result would be obtained if one took the limit
$E_n\rightarrow E_m$ in Eq.(\ref{impart_diff}).  We can now split the
double sum over $m, n$ in Eq.(\ref{isovec_1_real}) in two parts containing,
respectively, terms with $E_m \neq E_n$ and $E_m = E_n$ and perform the
integral over $\omega$ using Eqs.(\ref{impart_diff}) and
(\ref{impart_same}). One obtains
\be
\lefteqn{
[u(x)-d(x)]^{(1)} } \nonumber \\
&=& (2T^3) \frac{N_cM_N}{24I}
\sum\limits_{\scriptstyle n, m \atop {\scriptstyle {E_m \neq E_n}}}
\frac{1}{E_n - E_m}\langle n|\tau ^a|m\rangle
\nonumber \\
&& \times \langle m|\tau ^a(1+\gamma ^0\gamma ^3)\left\{ \eta (E_n)\delta
(E_n+P^3-xM_N)-\eta (E_m)\delta (E_m+P^3-xM_N)\right\} |n\rangle
\nonumber \\
&+& (2T^3)\frac{N_cM_N}{24I}
\sum\limits_{\scriptstyle n, m \atop {\scriptstyle {E_m = E_n}}}
\langle n|\tau^a|m\rangle
\langle m|\tau ^a(1+\gamma ^0\gamma ^3)\eta (E_n)
\delta^{\prime}(E_n+P^3-xM_N)|n\rangle .   
\nonumber \\
\label{isovec_1_discrete}
\ee
The double sum over levels in the first term here can be further simplified
making use of the identity
\be
&& \langle n|\tau ^a| m\rangle
\langle m|\tau ^a (1+\gamma ^0\gamma ^3) 
\delta (E_n + P^3 - x M_N) | n \rangle
\nonumber \\
&=& \langle m|\tau ^a| n\rangle
\langle n|\tau ^a (1+\gamma ^0\gamma ^3) 
\delta (E_n + P^3 - x M_N) | m \rangle ,
\ee
which holds for any pair of levels $m, n$.
\par
Equation (\ref{isovec_1_discrete}) pertains to a quasi--discrete spectrum;
in the continuum case one should replace there the summation over levels by
an integral over the continuous energy, omit the sum over terms with 
$E_m = E_n$ in the last line, and understand the poles $(E_n - E_m)^{-1}$ 
in the principal value sense.
\par
In Eq.(\ref{isovec_1_discrete}) the sum over $n$ runs over all quark
single--particle levels, both occupied and non-occupied. In order to
convert it to a more standard form, particularly for use in the numerical
calculations, we would like to rewrite the expression for the distribution
function as a sum over either occupied or non-occupied states ({\it cf.}\
the standard expression for the moment of inertia, Eq.(\ref{I})). To
achieve this we note that the following sum over {\it all} levels is zero:
\be
&& 2 \sum\limits_{\scriptstyle n, m\atop {\scriptstyle {E_m \neq E_n}}}
\frac{1}{E_n - E_m}
\langle n|\tau ^a|m\rangle
\langle m|\tau ^a (1+\gamma ^0\gamma ^3) \delta (E_n+P^3-xM_N) |n\rangle
\nonumber \\
&&+ \sum\limits_{\scriptstyle n, m\atop {\scriptstyle {E_m = E_n}}}
\langle n|\tau^a|m\rangle
\langle m|\tau ^a(1+\gamma ^0\gamma ^3) \delta ^{\prime
}(E_n+P^3-xM_N)|n\rangle   
\;\;\; = \;\;\; 0 .
\label{identity}
\ee
To prove this identity we note that the L.H.S.\ can equivalently be written
as the variation of a functional trace with a particular modified
Hamiltonian
\be
&& \mbox{Im} \; \int \frac{d\omega}{2\pi} \int\limits_{-\infty}^\infty dz^0  
e^{i (\omega - xM_N) z^0} \nonumber \\
&& \times \frac{\partial}{\partial \Lambda^a} \;
\mbox{Sp} \left[ \tau^a (1 + \gamma^0\gamma^3)
\exp ( i P^3 z^0 ) \frac{1}{\omega - H(U_c) - \Lambda^b \tau^b + i0}
\right] \, \Biggr|_{\Lambda = 0}\nonumber \\
&=& 
\int\limits_{-\infty}^\infty dz^0  e^{-i x M_N z^0} 
\; \frac{\partial}{\partial \Lambda^a} \;
\mbox{Sp} \left[ \tau^a (1 + \gamma^0\gamma^3)
\exp ( i P^3 z^0 ) \exp [ i z^0 (H(U_c ) + \Lambda^b \tau^b )] 
\right] \, \Biggr|_{\Lambda = 0} .
\nonumber \\
\label{modified_trace}
\ee
Due to the locality of the modified Hamiltonian, 
$H(U_c ) + \Lambda^b\tau^b$, we have
\be
\langle {\bf X} | \exp ( i P^3 z^0 ) \exp [ i z^0 
(H(U_c) + \Lambda^b \tau^b )] 
| {\bf X} \rangle &=& 0 \hspace{1cm} \mbox{if} \hspace{.5cm} 
|z^3| > z^0 .
\label{locality}
\ee
Extrapolating Eq.(\ref{locality}) to $z^3 \rightarrow z^0$ ({\it cf.}\ the
corresponding discussion in Refs.\cite{DPPPW96,DPPPW97}) we find that
Eq.(\ref{modified_trace}) is zero, which proves Eq.(\ref{identity}). Using
Eq.(\ref{identity}), and keeping in mind the definition of $\eta (E_{n})$,
Eq.(\ref{eta_def}), we may now easily obtain from
Eq.(\ref{isovec_1_discrete}) a representation of the distribution function
as a sum in which $n$ runs either only over occupied states,
Eq.(\ref{isovec_1_occ}), or non-occupied states,
Eq.(\ref{isovec_1_nonocc}). We note that the equivalence of the
representations of the distribution function as sums over occupied and
non-occupied states is confirmed also by the numerical calculations.
\par
When invoking the locality condition, Eq.(\ref{locality}), we have assumed
that the ultraviolet regularization of the theory does not violate this
property. For the regularization by Pauli--Villars subtraction,
Eq.(\ref{PV}), this is indeed the case. Regularization by a cutoff, on the
other hand, would violate Eq.(\ref{locality}), which amounts to violating
the anticommutation relation of the quark fields at space--like
separations, see the discussion in Refs.\cite{DPPPW96,DPPPW97}.

\newpage
%
%
\newpage
\newpage
\newpage
\newpage
\end{document}